\documentclass[proof]{pasj00}
\SetRunningHead{H. Uchiyama et al.}
{ No X-ray from the HESS J1741$-$302 region except a new
intermediate polar candidate }

\begin{document}
\title{No X-Ray Excess from the HESS J1741$-$302 Region\\ except a New Intermediate Polar Candidate }
\author{Hideki \textsc{Uchiyama},\altaffilmark{1}  Katsuji \textsc{Koyama},\altaffilmark{2}  Hironori \textsc{Matsumoto},\altaffilmark{3} 
Omar \textsc{Tibolla},\altaffilmark{4,5}  Sarah \textsc{Kaufmann},\altaffilmark{6} \\ and Stefan \textsc{Wagner}\altaffilmark{6} }
\altaffiltext{1}{Department of Physics, School of Science, The University of Tokyo, 7-3-1 Hongo, Bunkyo-ku, Tokyo 113-0033}
\email{uchiyama@juno.phys.s.u-tokyo.ac.jp}
\altaffiltext{2}{Department of Physics, Graduate school of Science, Kyoto University, Oiwake-cho, Kitashirakawa, Kyoto 606-8502}
\email{koyama@cr.scphy.kyoto-u.ac.jp}
\altaffiltext{3}{Kobayashi-Maskawa Institute for the Origin of Particles and the Universe, Nagoya University, Furo-cho, Chikusa-ku, Nagoya, 464-8601 }
\altaffiltext{4}{Max-Planck-Institut f\"ur Kernphysik, P.O. Box 103980, D96029 Heidelberg, Germany}
\altaffiltext{5}{Universit\"at W\"urzburg, 97074 W\"urzburg, Germany}
\altaffiltext{6}{Landessternwarte, Universit\"at Heidelberg, K\"onigstuhl, D 69117 Heidelberg, Germany}

\maketitle

\begin{abstract}
With the Suzaku satellite, we observed an unidentified TeV gamma-ray source HESS J1741$-$302 
and its surroundings.
No diffuse or point-like X-ray sources are detected from the bright southern emission peak of HESS J1741$-$302.
From its neighborhood, we  found a new intermediate polar candidate at the position 
of $(\alpha, \delta)_{\rm J2000.0} = (\timeform{17h40m35.6s}, \timeform{-30D14m16s})$, 
which is designated as Suzaku J174035.6$-$301416.
The spectrum of Suzaku J174035.6$-$301416  exhibits  emission lines at the 
energy of 6.4, 6.7 and 7.0 keV, which can be assigned as the K$\alpha$ lines from 
neutral, He-like and H-like iron, respectively. 
A coherent pulsation is found at a period of 432.1 $\pm$ 0.1 s. The pulse profile is quasi-sinusoidal 
in the hard X-ray band (4$-$8 keV), but is more complicated in the soft X-ray band (1$-$3 keV). 
The moderate period of pulsation, the energy flux, and the presence of the
iron K$\alpha$ lines indicate that Suzaku J174035.6$-$301416  
is likely an intermediate polar, a subclass of
 magnetized white dwarf  binaries (cataclysmic variables). 
 Based on these discoveries, we give some implications on the 
origin of GCDX and brief comments on HESS J1741$-$302 and PSR B1737$-$30.
\end{abstract}

\section{Introduction}
The most characteristic feature of the Galactic center (GC) region 
in the high energy band is the Galactic center diffuse X-ray emission (GCDX). 
It has strong emission lines at 6.4, 6.7 and 7.0 keV which are
K$\alpha$ lines from neutral, He-like and H-like Fe ions, respectively
 (e.g. \cite{Ko07b}).
Chandra \citep{Mu03,Mu09} resolved the GCDX into many faint
 X-ray sources. The integrated spectra of the point sources resemble the 
GCDX in the K$\alpha$ line
features (\cite{Mu04}, \cite{Ko07b}). 
The most  probable candidates responsible for the K$\alpha$ lines
are magnetic cataclysmic variables (mCVs) and/or active stars \citep{Mu04,Re09}.   
Therefore, a significant fraction of the GCDX, if not all, is due to 
these X-ray point sources. 
In fact, Chandra,  XMM-Newton and Suzaku have already found several point sources
which exhibit characteristic properties of mCVs such as moderate
pulse  periods and X-ray spectra with the prominent iron K$\alpha$ lines
(e.g. \cite{No09}).   
The equivalent widths of  the K$\alpha$ lines are smaller than 
those in the GCDX, and hence the origin of the GCDX is still debatable.

The H.E.S.S. Cherenkov telescope revealed that TeV gamma-ray sources \citep{Ah06a} and 
the large-scale diffuse TeV gamma-ray emission \citep{Ah06b} are also
present in the GC region.
The most plausible scenario
of  the origin of the TeV gamma-rays is interaction of molecular cloud with 
TeV energy protons (high-energy cosmic rays).  The cosmic ray must 
contain a large number of lower-energy proton, which may produce
a K$\alpha$  line of neutral iron (the 6.4 keV line) by the inner-shell ionization 
of the iron atoms in the molecular cloud.
Accordingly, it is conceivable that the TeV gamma-ray emission is associated with the 6.4 keV 
line. In fact, \citet{Ba09} reported possible association of TeV gamma-ray source HESS
J1745$-$303 with the 6.4 keV line.

HESS J1741$-$302 \citep{Ti09a, Ti09b} is one of the faintest unidentified TeV gamma-ray sources. It is
located near a relatively powerful 
radio pulsar PSR B1737$-$30 with the spin-down luminosity of $\sim 8 \times 10^{33}$ erg s$^{-1}$ \citep{Fo97}.
In order to search for possible connection between the TeV gamma-ray emission 
and the 6.4 keV line, we observed the region of HESS J1741$-$302  
with the Suzaku satellite twice.  
We found no X-ray excess from the bright southern emission peak of HESS J1741$-$302 nor PSR B1737$-$30.  
Instead, we discovered  a new probable mCV at the close vicinity of PSR B1737$-$30. 

This paper focuses on the discovery and property of the new mCV candidate, and
some implications on the origin of GCDX.  We also give brief comments on
the non-detection of X-ray from HESS J1741$-$302 and  PSR B1737$-$30.      

\section{Observations and Data Reduction}

\begin{table*}
  \caption{Suzaku observation data list.}
  \label{tab:obs_data}
  \begin{center}
    \begin{tabular}{lcccc}
      \hline\hline
      Target & Obs. ID & Start Time (UT) & Stop Time (UT) & Good Exposure Time (ks)*  \\
      \hline
      GC LARGEPROJECT15 & 503021010 & 2008-10-04 03:44:03 & 2008-10-05 10:57:24 & 50.1 \\
      HESSJ1741-B & 503077010 & 2009-02-26 01:00:60 &2009-02-27 11:35:19 & 46.2 \\
      \hline
      \multicolumn{4}{c}{* After the data screening described in the text.}
    \end{tabular}
  \end{center}
\end{table*}

The Suzaku X-ray Imaging Spectrometer (XIS) observed the regions near 
the radio pulsar PSR B1737$-$30  \citep{Fo97}  and HESS J1741$-$302 \citep{Ti09a, Ti09b}
twice; one was as the Suzaku deep survey project
of the Galactic center region and the other was the pointing observation 
on the unidentified TeV gamma-ray source HESS J1741$-$302. 
Details about these observations are listed in table~\ref{tab:obs_data}.

The XIS consists of four sets of X-ray CCD camera systems (XIS~0, 1, 2, and 3) placed on 
the focal planes of four X-Ray Telescopes (XRT) aboard the  Suzaku satellite. XIS~0, 2, and 3 have 
front-illuminated (FI) CCDs, while XIS~1 has a back-illuminated (BI) CCD. One of the FI CCD 
cameras (XIS~2) has been out of function since November 2006, and hence the data of XIS~2 have not been used.
Detailed descriptions of the Suzaku satellite, the XRT, and the XIS can  be found in \citet{Mi07}, \citet{Se07}, and \citet{Ko07a}.

The XIS observations were made with the normal mode. 
The CCD integration time is 8 s in this mode.
We used the cleaned XIS event data distributed from 
DARTS\footnote{http://www.darts.isas.jaxa.jp/astro/suzaku.}.
These cleaned data were processed with the processing version 2.2.11.22 and 2.2.11.24 for the
first and second observations, respectively.
The difference of the processing versions has no significant effect on the results.

In these processes, the data during the epoch of low Earth 
elevation angles less than 5 degrees 
(ELV$<\timeform{5D}$), day Earth elevation angles less than 20 degrees 
(DYE{\_}ELV$<\timeform{20D}$), and within the South Atlantic Anomaly were removed.
The good exposure times are listed in table \ref{tab:obs_data}. 
Although the XIS CCDs were significantly degraded by on-orbit particle radiation,
the CCD performances were restored by the spaced-row charge injection 
technique \citep{Pr08, Uc09}. 
Using on board calibration sources, we confirmed the spectral resolutions at 5.9 keV were $\sim$150 and
$\sim$200 eV (FWHM) for the FI and BI CCDs, respectively.

Due to a star tracker (STT) problem, the Suzaku attitude during the 
second observation was not locked at the programmed position, and hence
drifted by $\sim \timeform{2'}$ in the direction of the right ascension  between 
the start and stop time of the second observation.
The second observation data therefore were not used for the detailed imaging analysis (section \ref{ch:image}) 
but were used for the spectral and timing analysis (section \ref{ch:spectrum} and \ref{ch:timing})
after the position correction as described in section  \ref{ch:spectrum}.

\section{Analysis and Results} % section 3
We analyzed the data using the software package HEASoft 6.5.1. 
In this paper, uncertainties are quoted at the 90\% confidence range unless otherwise stated. 

\subsection{X-ray Image} %3-1
\label{ch:image}

To make X-ray images, we used only the data of October 2008 when the STT operated correctly. 
To increase statistics, the data of XIS~0, 1 and 3 were merged. 
Since the intensity of the non-X-ray background (NXB) 
depends on the geomagnetic cut-off rigidity (COR) \citep{Ta08},
we obtained COR-sorted NXB images using {\tt xisnxbgen}.  
After subtracting the NXB image, we divided the X-ray image 
by a flat-field image to correct vignetting.
The flat-field image was made with {\tt xissim} \citep{Is07}.

Since Suzaku has non-negligible position errors \citep{Uc08}, 
we fine-tuned the coordinate of the XIS image using the positions of catalogued visible stars. 
Since X-rays from normal stars are generally soft, we made a 0.5$-$2.0 keV band image. 
In the X-ray image, there are three point sources which correspond to catalogued 
infrared and visible stars,  whose 
positional errors are less than $\timeform{0."1}$ \citep{Cu03,Ho98}. 
To match the positions of the three stars, we simply shifted and 
rotated the Suzaku coordinate without any stretch or shear of the original 
Suzaku image. The result is shown in figure 1.
 After this fine-tuning, the averaged  difference between the catalogued 
 stars and the Suzaku position  is $\timeform{9"}$. 
 We regard this value as the systematic error.  
 We then constructed the 1$-$9 keV band image from the first observation (figure 2), in which  
 the NXB was subtracted and the vignetting was corrected in the same way as figure 1.

No significant X-ray emission is found from the radio pulsar PSR B1737$-$30, 
but a bright new source is discovered at about $\timeform{90"}$ 
east of PSR B1737$-$30 with the fine-tuned position of $(\alpha,  \delta)_{2000} = 
(\timeform{17h40m35.6s}, \timeform{-30D14m16s})$. Here we designated this source as 
Suzaku~J174035.6$-$301416. The uncertainties are $\timeform{11"}$ and 
$\timeform{9"}$ due to statistical and systematic errors, respectively.  
Thus overall uncertainty is evaluated as  $\timeform{14"}$.

Near Suzaku~J174035.6-301416, we found two catalogued 
sources, SAX J1740.5-3013 \citep{Hu99}  and AX J1740.5-3014 \citep{Sa02}.
The offset from  Suzaku~J174035.6-301416 and error radius of  SAX J1740.5-3013 are 
  $\timeform{62"}$ and about $\timeform{50"}$, respectively,
while those of AX J1740.5-3014 are $\timeform{107"}$ and about $\timeform{90"}$.
With the error radius of Suzaku J174035.6-301416 of  $\timeform{14"}$,  the error regions of the three objects 
overlap with each others, and hence could be the same source.
These sources were also observed by \citet{Pa03} with the INTEGRAL satellite,
but were not detected due to the larger detection limit than that of Suzaku.

\begin{figure}[htpb] %figure 1
\begin{center}
\FigureFile(80mm,80mm){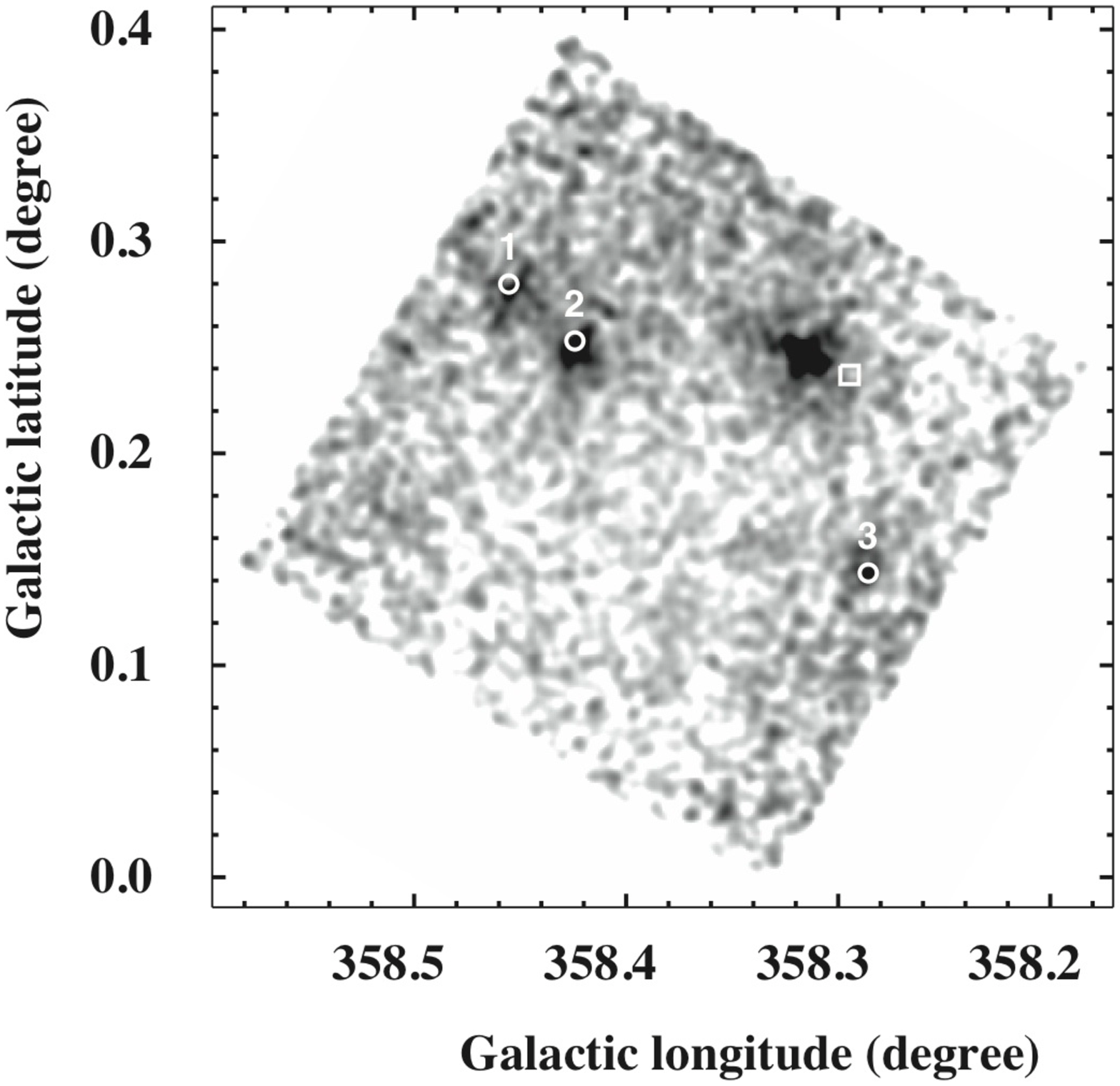}
\end{center}
\caption{X-ray image in the 0.5$-$2 keV band after the coordinate correction (see text in section \ref{ch:image}). The numbered circles represent catalogued stars used as references of the position. (1) DENIS-P J174047.6$-$300613 (2) HD 316162 (3) DENIS-P J174054.9$-$301915 \citep{Cu03,Ho98}. The rectangle mark shows the position of PSR B1737$-$30 \citep{Fo97}. The NXB was subtracted and the vignetting was corrected (see text). This image was smoothed using a Gaussian function with $\sigma=\timeform{0'.35}$.} 
\label{fig:Map1}
\end{figure}

\begin{figure} %figure 2
\begin{center}
\FigureFile(80mm,80mm){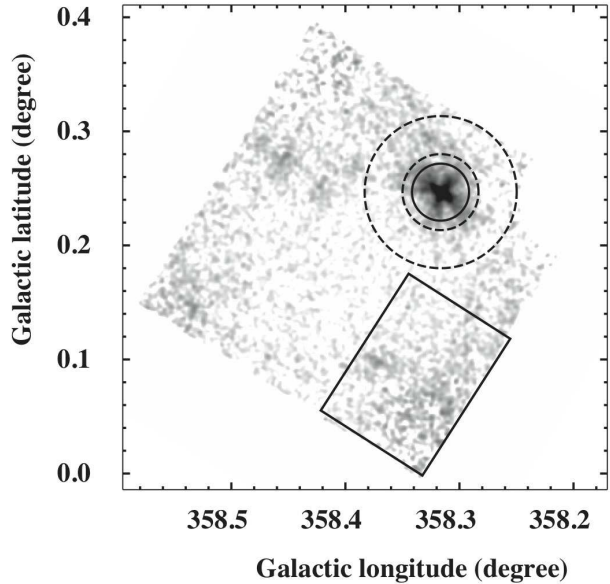}
\end{center}
\caption{X-ray image in the 1$-$9 keV band after the coordinate correction. The source and background regions for spectra and light curves of Suzaku~J174035.6$-$301416 are shown by the solid circle and dashed annulus, respectively. The solid box shows the source region of HESS ~J1741$-$302. 
The NXB was subtracted and the vignetting was corrected in the same way as figure 1. This image was smoothed using a Gaussian function with $\sigma=\timeform{0'.2}$.} 
\label{fig:Map1}
\end{figure}

\subsection {Spectrum of Suzaku J174035.6$-$301416 } %3-2
\label{ch:spectrum}

The spectra of Suzaku J174035.6$-$301416 
were extracted from the solid circle region with a radius of $\timeform{1.5'}$ and subtracted by
the background spectra evaluated in the near region described by the
dashed annulus with radii of $\timeform{2'}$ and  $\timeform{4'}$ shown in figure 2.

In the second observation, however,  the STT was troubled and the position
drifted slowly. We therefore divided the data of the second observation into six data sets. 
In the each data set, the attitude drift is less than $\timeform{20"}$, which 
is comparable to the  positional uncertainty of $\timeform{14"}$.
We made X-ray images from these six data sets respectively and 
decided the peak positions of Suzaku~J174035.6$-$301416  in the each data.
We corrected the coordinates of the six data sets to match the peak positions 
with each others. Then we made positional fine tuning with the same method as section \ref{ch:image}. 
After these corrections, the spectra and light curves (section 3.3) for 
the second observation were extracted.

The difference between the fluxes of the first and second observations are less than 5 \%
 and we did not find clear difference between the shapes of the spectra.  
We thus combined the spectra of XIS of the both observations to increase the statistics.  
The spectra of the two FIs (XIS 0, 3) were co-added because the response 
functions are almost the same between the FIs. 
The X-ray spectrum is shown in figure 3, where  only the spectrum of the FI CCDs is given for simplicity.
XIS response and XRT auxiliary files were made using 
{\tt xisrmfgen} and {\tt xissimarfgen} \citep{Is07} for the each observation.
These response and auxiliary files were also combined. 

\begin{figure} %figure 3
\begin{center}
\FigureFile(80mm,80mm){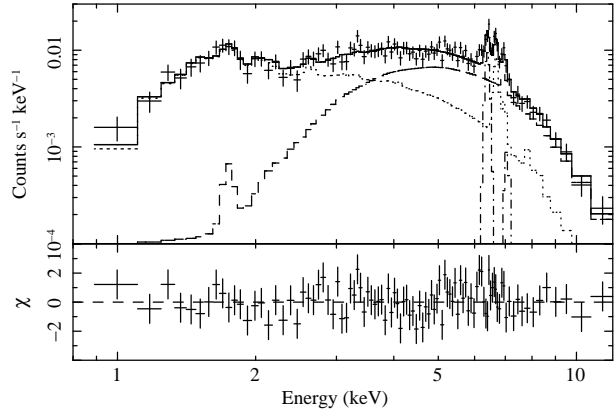}
\end{center}
\caption{Background-subtracted spectrum of Suzaku J174035.6$-$301416  for the FI CCDs.
The source spectrum was extracted from the solid circle  in figures 2, while
the local background spectrum was taken from the dashed 
annulus described in figure 2.  The vertical error bars of the each data point
are the 1$\sigma$ error. The solid line is the best-fit result of the two-components CIE model shown in table 2(b). 
The dashed and dotted lines represent the high- and low-temperature CIE components.
The dash-dot lines represent the neutral iron K$\alpha$ and $\beta$ lines.
Although the fitting was made simultaneously for the FI and BI spectra, only the FI result is given for simplicity }
\label{fig:Spec3}
\end{figure}

The X-ray spectra exhibit three lines at 6.4, 6.7, and 7.0 keV. 
In order to identify these lines, the spectra were fit with a phenomenological model of an absorbed 
power-law plus three narrow Gaussian lines in the 5$-$10 keV band.  
The cross section of the photoelectric absorption was obtained from \citet{Mo83}.
The best-fit line energies of these three lines are $6.40^{+0.04}_{-0.02}$ keV,  $6.67 \pm 0.02$ keV 
and $6.95^{+0.04}_{-0.05}$ keV, while the equivalent widths (EW) are 210$^{+30}_{-50}$~eV, 
190$^{+60}_{-40}$~eV  and 130$^{+40}_{-50}$ eV, respectively.  
From the line energies, these are identified as the K$\alpha$ lines from neutral, He-like and H-like iron.

We fitted the X-ray spectra of the FI and BI simultaneously  with a one-component collisional ionization 
equilibrium (CIE) plasma model (APEC: Smith et al. 2001).
We added  two Gaussians, Fe \emissiontype{I} K$\alpha$ (6.40~keV) and Fe \emissiontype{I} K$\beta$ 
(7.06 keV). 
The energy of Fe \emissiontype{I} K$\beta$ is near to Fe \emissiontype{XXVI} K$\alpha$, and hence
was not separable. In order to estimate the Fe \emissiontype{XXVI} K$\alpha$ emission
accurately, we included the Fe \emissiontype{I} K$\beta$  (7.06 keV). According to Kaastra \& Mewe (1993), 
the ratios of the center energies and line flux for the neutral iron K$\alpha$ and K$\beta$ 
lines were fixed to be 1:1.103 and 1:0.125, respectively. 
The abundance of the CIE plasma model 
was free but the relative ratio among elements was fixed to be the solar value \citep{An89}.
This model is rejected with $\chi^2$/d.o.f. = 363.3/166. 
The same model but two-components CIE plasma with the common absorption and abundance is also rejected with $\chi^2$/d.o.f. = 275.7/164.  

Since large residuals are found in the low energy band, we applied a one-component CIE plasma with partial absorption model.
This model gives a marginal $\chi^2$/d.o.f. of  204.9/164. The best-fit parameters are listed in table 2(a).  

We finally  tried a  two-components CIE plasma with the independent absorptions and common abundance model, and obtained
the most reasonable fit with $\chi^2$/d.o.f. = 190.6/163. The best-fit parameters and the result are shown in table 2(b) and figure 3.

\begin{table*} %table 2
\caption{Best-fit parameters of the model fittings for the background-subtracted spectra\footnotemark[$*$].}
\label{tab:partial}
\begin{center}
\begin{tabular}{lcc}
\multicolumn{3}{l}{(a) one-component CIE plasma with partial absorption model}\\
\hline\hline
\multicolumn{3}{l}{Model: Abs[1]$\times$\{(1-$\alpha$)+$\alpha \times $Abs[2])$\times$(APEC+Neutral Iron Lines)\}}\\
\hline
Iron Lines & Fe $\emissiontype{I}$ K$\alpha$ & Fe $\emissiontype{I}$ K$\beta$ \\
\hline
Energy (keV) & $6.40^{+0.03}_{-0.02}$ &7.06\footnotemark[$\dagger$] \\
Flux ($10^{-6}$ ph~s$^{-1}$cm$^{-2}$) & $8.8^{+1.7}_{-1.3}$ &1.1\footnotemark[$\dagger$]\\
\hline
Parameter & \multicolumn{2}{c}{APEC}\\
\hline
$kT$ (keV) & \multicolumn{2}{c}{$9.5^{+1.2}_{-0.9}$} \\
Abundance (solar) & \multicolumn{2}{c}{$0.49^{+0.11}_{-0.10}$} \\
Normalization\footnotemark[$\ddagger$] & \multicolumn{2}{c}{$2.8^{+0.1}_{-0.2}$}\\
\hline
Parameter & Abs[1] & Abs[2]\\
\hline
$N_{\rm H}~(10^{22}$~cm$^{-2}$) & $1.7_{-0.2}^{+0.1}$ & $16 \pm 3$ \\
Covering factor $\alpha$  & $-$& $0.71_{-0.03}^{+0.02}$\\
\hline
Flux ($10^{-12}$ erg s$^{-1}$ cm$^{-2}$)\footnotemark[$\S$]  & \multicolumn{2}{c}{2.0 (2.2)}\\
\hline
\multicolumn{3}{c}{$\chi^2$/d.o.f. = 204.9/164}\\
\hline
\\
\multicolumn{3}{l}{(b) two-components CIE plasma model}\\
\hline\hline
\multicolumn{3}{l}{Model: Abs[1]$\times$(APEC[1]+Neutral iron lines)+Abs[2]$\times$APEC[2] }\\
\hline
Iron Lines & Fe $\emissiontype{I}$ K$\alpha$ & Fe $\emissiontype{I}$ K$\beta$ \\
\hline
Energy (keV) & $6.40^{+0.03}_{-0.02}$ &7.06\footnotemark[$\dagger$] \\
Flux ($10^{-6}$ ph~s$^{-1}$cm$^{-2}$) & $8.4^{+1.7}_{-1.4}$ &1.0 \footnotemark[$\dagger$]\\
\hline
Parameter & APEC[1] & APEC[2]\\
\hline
$kT$ (keV) & $64 (\ge 44)$ & $6.0 \pm 1.1$ \\
Abundance (solar)\footnotemark[$\|$] & \multicolumn{2}{l}{~~~~~~~~~~~~~~~$1.3^{+0.2}_{-0.3}$} \\
Normalization\footnotemark[$\ddagger$]  &$1.6^{+0.2}_{-0.1}$ & $0.62_{-0.03}^{+0.04}$ \\
\hline
Parameter & Abs[1] & Abs[2] \\ 
\hline
$N_{\rm H}$ ($10^{22}$~cm$^{-2}$) & $11^{+2}_{-1}$ & $1.6 \pm 0.1 $  \\
\hline
Flux ($10^{-12}$ erg s$^{-1}$ cm$^{-2}$)\footnotemark[$\S$] & \multicolumn{2}{c}{2.1 (2.3)}\\
\hline
\multicolumn{3}{c}{$\chi^2$/d.o.f. = 190.6/163} \\
\hline
\\
\end{tabular}
\end{center}
\footnotemark[$*$] Errors show 90\% confidence range.\\
\footnotemark[$\dagger$] The energy and flux of Fe $\emissiontype{I}$ K$\beta$ line are fixed to be 1.103 and 0.125 times of those of Fe $\emissiontype{I}$ K$\alpha$ line, respectively.\\
\footnotemark[$\ddagger$] The units of $10^{-17}/(4\pi D^2)\int n_e n_H dV$, where $D$, $n_e$ and $n_H$ are the distance to the source (cm), 
   the electron density (cm$^{-3}$), and the hydrogen density (cm$^{-3}$), respectively.\\
\footnotemark[$\S$] Observed flux in the range of 2$-$10 keV. The values in parentheses are corrected with the absorption $N_{\rm H}$ = 1.7 $\times 10^{22}$~cm$^{-2}$ (a) or 1.6 $\times 10^{22}$~cm$^{-2}$ (b) as the interstellar absorptions.\\
\footnotemark[$\|$] The abundances are common between APEC[1] and APEC[2] in (b).\\
\end{table*}

\subsection {X-ray limits from PSR B1737$-$30 and HESS J1741$-$302} %3-3
Although the detailed morphology of HESS J1741$-$302 may not be
final \citep{Ti09a,Ti09b},  the global size and the peak position are reliable.
HESS J1741$-$302 consists of southern and northern bright hot spots, whose sizes 
are $\sim\timeform{0D.1}$ and  $\sim\timeform{0D.2}$ (FWHM)  \citep{Ti09b}. The former spot is   
corresponds to the south west corner of the Suzaku field (the lower corner in figure 2).
Compared to the other region in figure 2, we see no excess X-rays from this region in
the 1--9 keV band.  Also no X-rays are found from the position of PSR B1737$-$30
(the rectangle in figure 1).

In order to estimate the
upper limit of the fluxes, we extract the spectrum  from the southern 
TeV gamma-ray emission peak region of HESS J1741$-$302,  
the solid box ($\timeform{6'}\times \timeform{9'}$) in figure 2.
The background was taken from the remaining region of the XIS field 
excluding the $\timeform{4'}$-radius circle region around Suzaku~J174035.6$-$301416. 
The background subtracted spectrum 
was fit with an absorbed power-law.
The photon index was  fixed to 2.0, the typical value of pulsar wind nebula, 
while the absorption column was assumed to be the same 
as Suzaku~J174035.6$-$301416, $N_{\rm H} = 1.6 \times 10^{22}$ cm$^{-2}$.   
Then the flux upper limit  
was evaluated to be $< 1.6 \times 10^{-13}$ erg~s$^{-1}$~cm$^{-2}$  in the 90\% confidence level.

With the same method, we estimated the X-ray upper-limit
of PSR B1737$-$30  in the $\timeform{1'}$-radius 
circle excluding the $\timeform{1'.5}$-radius circle region around Suzaku~J174035.6$-$301416.
 For the background, we used the same background region above. 
 Since  the half  power diameter of the XRT is $\sim \timeform{2'}$, the spectrum  from the region including 
PSR B1737$-30$  is 
 contaminated by Suzaku~J174035.6$-$301416.
Using  the XIS simulator  {\tt xissim} \citep{Is07}, we calculated 
the contaminating flux to be 16\% of  Suzaku~J174035.6$-$301416. 
After subtraction of this contamination, the spectrum from the region including 
PSR B1737$-30$ was fitted with the same model above.
Then we evaluated the flux upper limit  to be $< 3.5 \times10^{-13}$ erg~s$^{-1}$~cm$^{-2}$ in the 90\% confidence level.

The flux upper limit from the former large region $(\timeform{6'} \times \timeform{9'})$ is smaller 
than that for the region including PSR B1737$-$30, which is assumed to 
be a point source. It seems a little strange. The later region is, however, small 
and the photon statics are low. On the other hand, the contamination from 
Suzaku~J174035.6$-$301416 is quite high. Thus the upper limit becomes large.

\subsection{Timing analysis of Suzaku~J174035.6$-$301416 }
\label{ch:timing}

We  searched for a coherent pulsation in the 1$-$9~keV band from the FI data of the both observations.
We did not use the BI data because it had high and time-variable NXB, especially above the 7 keV band. 
The fast Fourier transform (FFT) analysis revealed a clear peak at $\sim 2.3\times10^{-3}$ Hz as is shown in figure 4.
We then searched for an accurate pulse period with the folding technique, and found a pulse period of $432.1 \pm 0.1$ s. 
The error of the pulse period was estimated with the method given in \citet{La96}.

\begin{figure} %figure 4
\begin{center}
\FigureFile(80mm,50mm){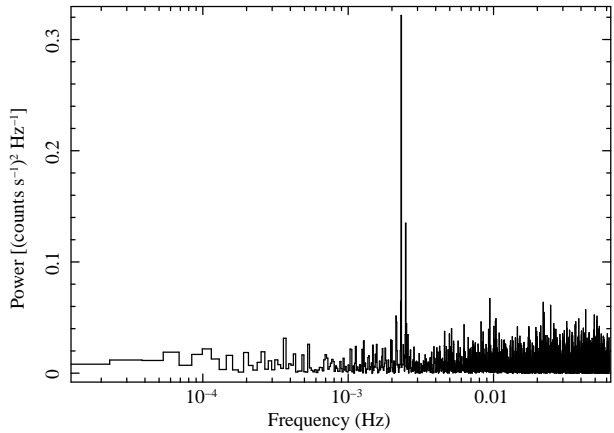}
\end{center}
\caption{Power spectrum (FFT) in the 1$-$9~keV band.}
\label{fig:Time5}
\end{figure}

\begin{figure} %figure 5
\begin{center}
\FigureFile(80mm,50mm){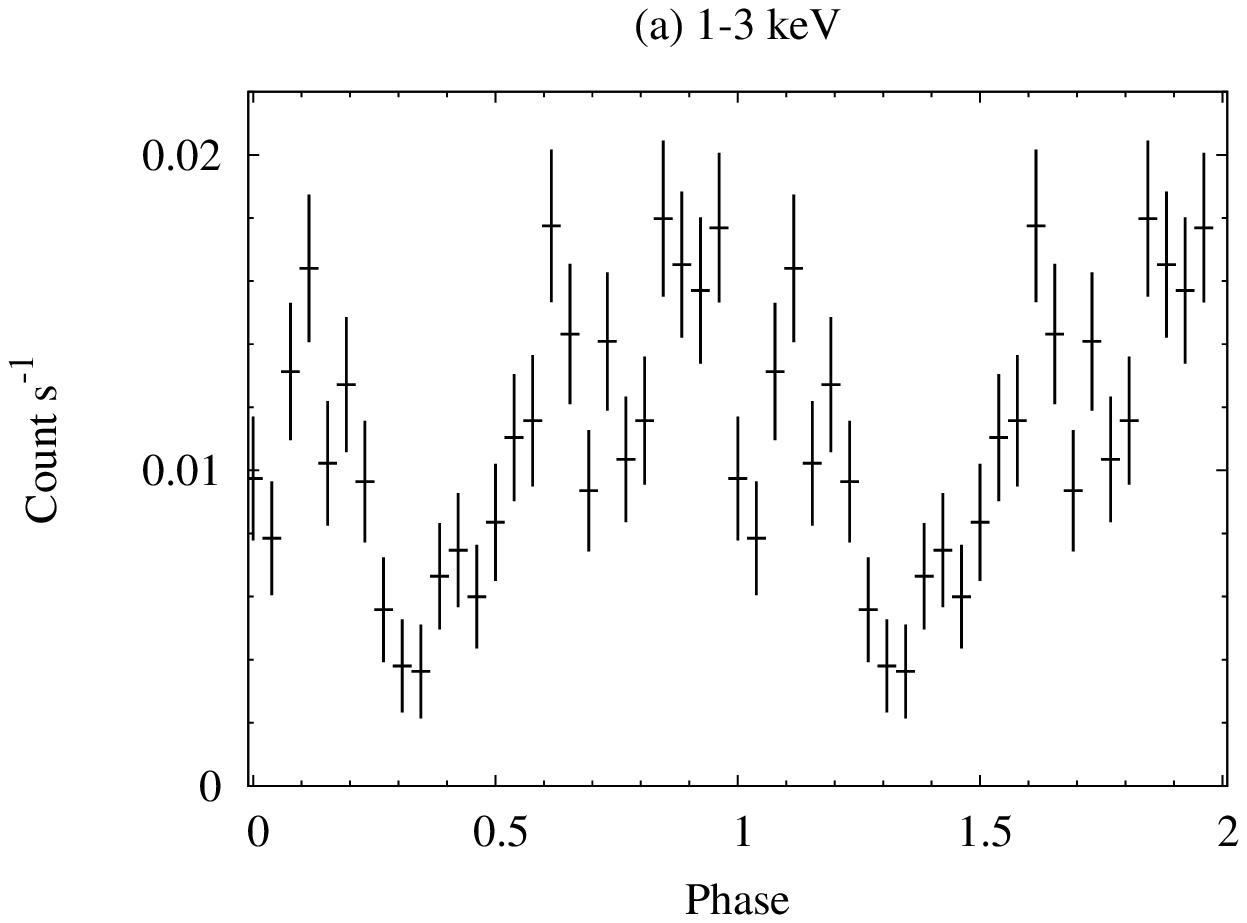}\\
~\\
\FigureFile(80mm,50mm){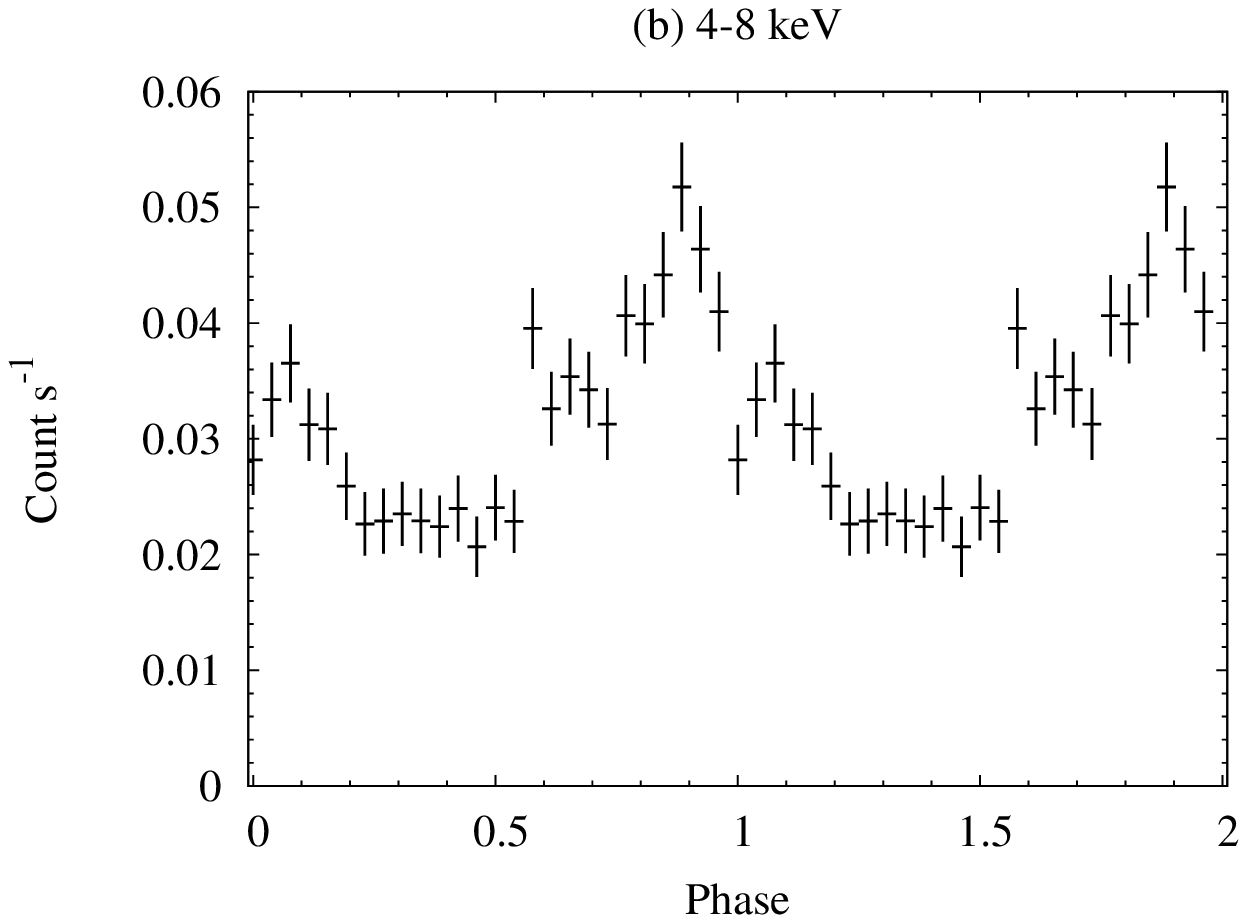}\\ 
~\\
\FigureFile(80mm,50mm){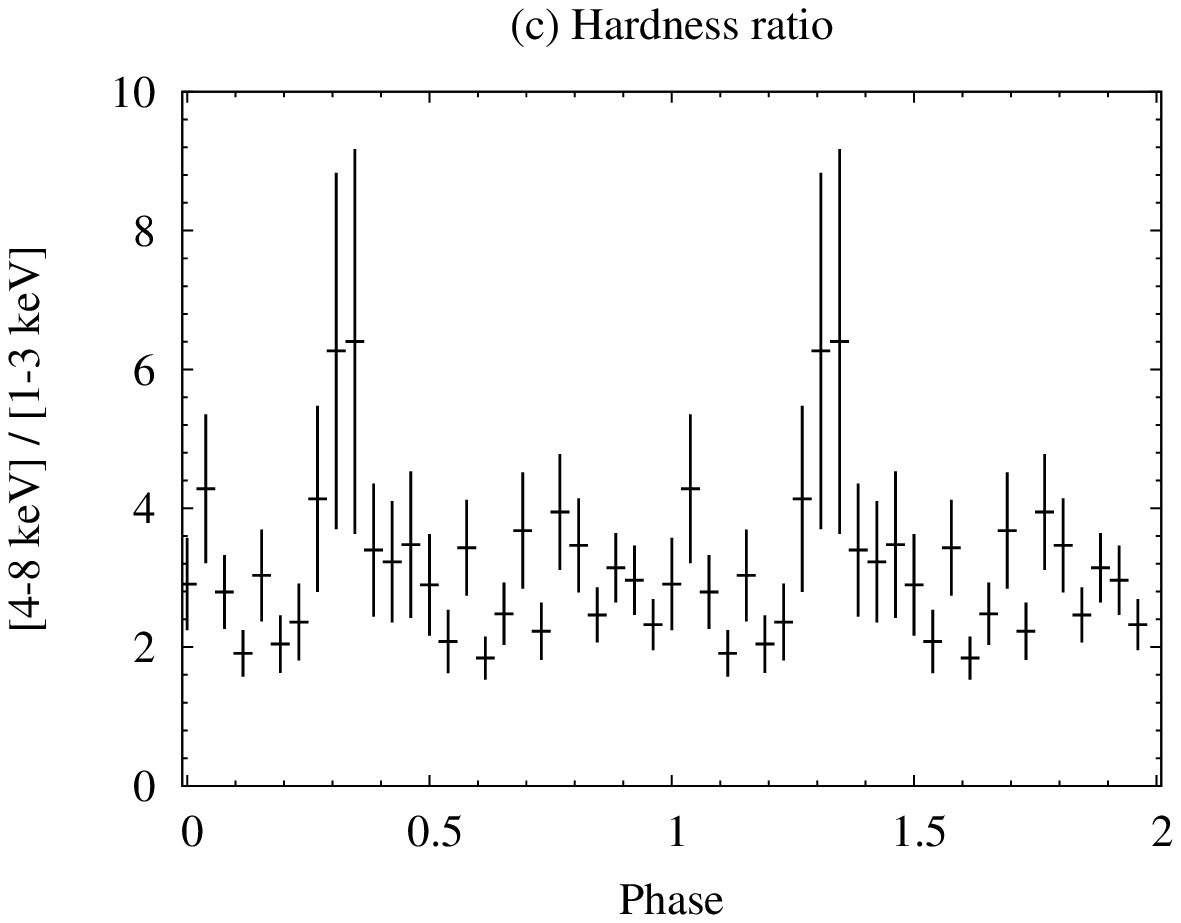}
\end{center}
\caption{(a) Pulse profile in the 1$-$3 keV band, (b) that of the 4$-$8 keV band. The count rates of the background region are subtracted. (c) Hardness ratio of the 4$-$8 keV to 1$-$3 keV. Folded period is 432.1 s. The vertical error bars of each data point are the 1$\sigma$ error.}
\label{fig:Time5}
\end{figure}

The sub-peak at $2.5\times10^{-3}$ Hz in figure 4 would be the beat frequency between the 432.1-s period  and  the Suzaku orbital period of about 96 minutes. 

Since we found the two-components CIE plasma in the spectral analysis, we
made folded pulse profiles in the 1$-$3 keV and 4$-$8 keV bands representing the low- and high-temperature components, respectively. 
These folded pulse profiles are shown in figure 5, where normalized count rates of the background region, 
$4.7 \times 10^{-3}$ (1--3 keV) and $6.2 \times 10^{-3}$ (4--8 keV) counts s$^{-1}$, were subtracted.
We also show the hardness ratio between these bands.
The pulse profiles appear to be different between the two components.
The pulse profile of the high-temperature plasma is quasi-sinusoidal with small humps 
in the both sides of the main peak, while that of the low-temperature plasma is more complicated with nearly equal three peaks.  
The minima are deeper, but narrower, in the soft band compared to the
hard band. The former is expected if $N_{\rm H}$ is varying as is common in intermediate polars (IPs)  \citep {No89}.

%%\subsection{Phase divided Spectral analysis of Suzaku~J174035.6$-$301416 }

\section {Discussion}

\subsection{Nature of  Suzaku~J174035.6$-$301416 }

The fine-tuned position of Suzaku J174035.6$-$301416  at $(\alpha,  \delta)_{2000} = 
 (\timeform{17h40m35.6s}, \timeform{-30D14m16s})$ is different from that of  PSR B1737$-$30 
 at $(\alpha, \delta)_{2000} = (\timeform{17h40m33.82s}, \timeform{-30D15m43.5s})$ \citep{Fo97}. 
The X-ray spectra of Suzaku J174035.6$-$301416  exhibit  K$\alpha$ lines at 6.4, 6.7, and 7.0 keV  
with the best-fit EW values of $\sim$210, $\sim$190, and $\sim$130~eV, respectively. 
These  EW values are  nearly the same as the mean EW 
of mCVs (\cite{Ez99}), which are $\sim$100, $\sim200$, and $\sim$100~eV.
We found a coherent and clear pulsation of 432.1 s from Suzaku J174035.6$-$301416. 
This is usual spin period of mCVs , which typically ranges 
from 30 s to $2 \times 10^4$ s \citep{Sc10}.
Although we found no periodicity due to orbital 
modulation of Suzaku J174035.6$-$301416, this 432.1-s period would be due to spin, 
because this value is smaller than typical orbital period, which ranges 
from 2 to 10 hours  \citep{Sc10}.

As shown in section 3.1,  Suzaku~J174035.6$-$301416  is likely to be identical to 
AX~J1740.5$-$3014 and SAX~J1740.5$-$3013.  The latter two sources were 
observed in September 1995 and in April 1998, respectively.  
The data qualities of these sources , however, were  too limited to 
perform spectral and timing analysis.  Nevertheless, \citet{Sa02} and 
\citet{Hu02} reported the fluxes  of these sources are  2.5, and 2.4 $\times 10^{-12}$ erg s$^{-1}$ cm$^{-2}$, respectively.  
Since the Suzaku flux is 2.1 $\times 10^{-12}$ erg s$^{-1}$ cm$^{-2}$, we see no large variability in the long time span of $\sim 10$ years.

The overall spectrum of Suzaku J174035.6$-$30141 was well fit with the two-components 
CIE plasma with temperatures of 6.0 and 64 keV or marginally fit with the one-component 
CIE plasma with partial covering. The both models require the 
K$\alpha$ (6.40~keV)  from neutral iron. 
Ezuka and Ishida (1999) also reported that the spectra of mCVs can be described by 
a thin thermal plasma model
with a mean temperature of $\sim$20 keV plus 6.4 keV line. 
The one-component CIE plasma with partial covering model
gives a temperature of 10 keV, which resembles to the general feature of mCVs.
We however favor the two-components CIE plasma model, because $\chi^2$ 
is better and the energy dependent pulse profile is difficult to explain with 
the one-component CIE plasma model.

The slow pulsation and X-ray spectrum, in particular the iron K$\alpha$ line 
features suggest that Suzaku J174035.6$-$301416  is a new persistent mCV. 
It is more likely to be an IP and  not a polar, because the 432.1-s period 
is too short as a synchronized orbital-spin period 
($\ge$ 4000 s for all the catalogued polars by Ritter \& Kolb 2003).
The lack of long term variability in the X-ray flux of this IP
candidate is indeed typical of this class, as they do not show off states, and 
only have very short dwarf nova-type outbursts.
For the definitive classification,  the optical spectrum and light curve of  
Suzaku J174035.6$-$301416 are required. Using the SIMBAD database\footnote{http://simbad.u-strasbg.fr/simbad/}, 
we searched a corresponding object but there is no catalogued optical object within $\timeform{80"}$ 
around  Suzaku J174035.6$-$301416. 

We suggest  that a large absorption for the 64 keV plasma  
is due to the circum-stellar gas, which could be 
up to a few $\times 10^{23}$~cm$^{-2}$ \citep{Ez99}. 
This large amount of circum-stellar gas can naturally explain the origin of the strong 
6.4 keV line from neutral iron (see e.g. \cite{Ez99}).

The spectrum of the standard model with a cylindrical emission region is given by multi-temperature 
plasma components with power-law emission measure distribution of index $\sim -0.5$ \citep{Is94}.
The multi-components plasma can be approximated by two representative temperatures. 
The ratio of the volume emission measures of the 6.0 keV plasma to the 64 keV plasma is 1:2.6.
This ratio does not follow the power-law relation of $\sim -0.5$. 
It is also strange that the 6.0 keV plasma emission does not suffer the large 
absorption, if that is attributable to the circum-stellar gas. 
This new IP seems difficult to be explained with  the standard emission model and geometry.
However, the spectra of IPs are complex with multi-temperatures, soft 
blackbody components (e.g. \cite{Ev07}) and hard reflection components 
 (e.g. \cite{Re04}) so on. In addition, the spectra depend strongly on spin and orbital phase 
 (e.g. \cite{Al98}).  Our spectral model is unlikely to be a unique description. 
The statistic of this source is, however, not so high, and the detail modeling is beyond the 
scope of this paper.

The absorption for the low temperature component, on the other hand, can be 
interstellar medium to Suzaku~J174035.6$-$301416.
The typical absorption to the GC region is 
$N_{\rm H}=6\times10^{22}$~cm$^{-2}$~(e.g. \cite{Sa02},  \cite{Ry09}). 
However this
value rapidly decreases as one moves away from the Galactic Plane. 
For example, G359.1$-$0.5, a GC supernova remnant located  $\timeform{0D.3}$--$\timeform{0D.5}$ away 
from the plane, has absorption of about  $N_{\rm H}=2\times10^{22}$~cm$^{-2}$ \citep{Oh10}.  
Suzaku~J174035.6$-$301416  has $N_{\rm H}=1.6\times10^{22}$~cm$^{-2}$, 
nearly identical to G359.1$-$0.5, and hence might be a GC source.
Assuming the distance to GC as  8~kpc (e.g. \cite{Ma09}),  the absorption-corrected
flux of $2.3 \times10^{-12}$~erg~cm$^{-2}$~s$^{-1}$ in the
2--10 keV band is converted to the source luminosity of 
$1.8 \times 10^{34}$~erg~s$^{-1}$ in the 2--10~keV band, 
which places Suzaku~J174035.6$-$301416 
as the brightest class of the IP \citep{Pa94}, and 
nearly the same order of faint neutron star binaries.  
However neutron star binaries do not exhibit strong Fe K$\alpha$  line emissions 
and the EWs are typically less than 100 eV \citep{Ng10}.  
Suzaku~J174035.6$-$301416 might be one of  the most powerful mCVs  in the GC region.
IPs are, however, well known to be intrinsically absorbed (e.g. Evans \& Hellier 2007), 
and thus the measured absorption is not clear to be related to the interstellar absorption.
Our estimation of the distance and luminosity  should have large uncertainty.
If we assume the luminosity of  typical IPs, $10^{31}$--$10^{33}$ erg s$^{-1}$ \citep{Ez99}, 
as that of  Suzaku~J174035.6$-$301416,
the distance estimated  to be 0.2--2~kpc from the observed flux.

\subsection{Comments on the origin of HESS J1741$-$302 and the GCDX}

We find no excess X-rays from the southern emission peak region of HESS J1741$-$302 nor from  the
nearby radio pulsar PSR B1737$-$30. 
We constrain the flux ratio of TeV gamma to hard X-ray ($F_{\rm 1-10 TeV}/F_{\rm 2-10 keV}$) to larger than $\sim 12$.  
\citet{Mam09} reported that X-ray emission is detected from the northern emission peak of HESS J1741$-$302 and $F_{\rm 1-10 TeV}/F_{\rm 2-10 keV}$ is $\sim 6$.
These large ratios may suggest that HESS J1741$-$302 is probably  a "dark accelerator" like HESS J1616$-$508 \citep{Ma07}.   
Unlike HESS J1745$-$303 \citep{Ba09}, we  find no hint of the 6.4 keV line from the diffuse region of 
HESS J1741$-$30 or  no possible correlation between the GCDX and TeV gamma-ray 
emission, at least for HESS J1741$-$302.

On the other hand, our discovery of an IP candidate may give new insight for the origin of GCDX. 
As we already note, significant fraction of the iron K$\alpha$ lines in GCDX, would be  due to mCVs.  
However the mean equivalent widths of the 6.4 keV and 6.7 keV lines in the GCDX are 
$\sim400$~eV and $\sim450$~eV, respectively (e.g. \cite{Ko09}). 
These are significantly larger than those of  Suzaku~J174035.6$-$301416, and
 the other bright mCVs (\cite{Ez99}). 
 This statement may be true for the most of the mCVs, which are 
 bright enough to measure the iron lines.  
To solve this entangled issue, we need to measure the flux of iron K$\alpha$ 
lines from  fainter  mCV and/or the other active stars.

\bigskip

The authors thank all of the Suzaku team members, especially T. Tsuru, 
M. Nobukawa, K. Makishima, T. Yuasa and M. Ishida for their comments and useful information on the XIS performance.
This work is supported by the Grant-in-Aid for the Global COE Program "The Next Generation 
of Physics, Spun from Universality and Emergence" and Challenging Exploratory Research (KK) 
from the Ministry of Education, Culture, Sports, Science and Technology (MEXT) of Japan.
HU is supported by JSPS Research Fellowship for Young Scientists.

\end{document}